\documentclass{aa}

\usepackage{epsfig,epsf}
\usepackage{txfonts,natbib}
\usepackage{amssymb}
\usepackage{latexsym}
\usepackage{graphicx}
\usepackage{natbib}

\newcommand{\vek}[1]{{\mbox {\boldmath $#1$}}}
\newcommand{\vekt}[1]{{\mbox {{\rm \bfseries #1}}}}
\begin{document}
\title{Massive galaxy clusters as gravitational telescopes for distant
supernovae} 
\author{Christofer Gunnarsson \and Ariel Goobar}
\offprints{C.~Gunnarsson, \email{cg@physto.se}}
\institute{Department of Physics, Stockholm University, AlbaNova,
SE-106 91 Stockholm, Sweden} 
\date{Received / Accepted}
\abstract{We investigate the potential of using massive clusters as
gravitational telescopes for searches and studies of supernovae of
Type Ia and Type II  
in optical and near-infrared bands at central wavelengths in the interval
0.8-1.25 $\mu$m.  
Using high-redshift supernova rates derived from the measured star
formation rate,   
we find the most interesting effects for the detection of 
core-collapse SNe in    
searches at limiting magnitudes  $m_{\rm lim}\sim 25-26.5$ mag, where
the total detection rate could be significantly enhanced and the
number of detectable events is 
considerable even in a small field. For shallower searches, $\sim$24, 
a net gain factor of up to 3 in the discovery rate could be obtained, and yet 
a much larger factor for very high source redshifts. For programs
such as the GOODS/ACS transient survey, the discovery rate of supernovae 
beyond $z\sim 2$ could be significantly increased if the observations were done
in the direction of massive clusters.
For extremely deep observations, $m_{\rm lim} > 27$ mag, or for very
bright SNe (e.g.~Type Ia) the  
competing effect of field reduction by lensing dominates, and fewer supernovae
are likely to be discovered behind foreground clusters.  

}  

\maketitle
\section{Introduction}
Massive galaxy clusters, used as gravitational telescopes (GTs), are
extremely useful tools for the studies of faint high redshift galaxies
in wavelengths ranging from optical to submillimetre as demonstrated
e.g.~in \citet{ellis2001,hu2002,lemoine2002,biviano,smail}.  In this
work we explore the potential of GTs for magnifying high redshift
supernovae (SNe) in optical and near-infrared (NIR) wavelengths,
thereby increasing the chances of detection.  A competing effect
related to the use of GTs is due to the spreading of the field by the
lens, analogous to what happens when looking through a magnifying
glass: a smaller, although magnified, portion of the field is actually
observed. The effect is sometimes referred to as \emph{amplification
bias}.  For any specific field-of-view (FOV) it is not obvious a
priori which of the two effects dominates when looking for distant
supernovae.  The net gain depends upon 1) the lens and source parameters:
the mass distribution of the lensing cluster, the intrinsic rate and brightness for core
collapse and Type Ia supernovae as a function of redshift. 2) The observational
set-up: the limiting search magnitude and the choice of wavelength band.  
In this
paper we consider several scenarios relevant for supernova searches.

Studying supernova (SN) rates at the highest possible redshifts provides
critical information for 
the understanding of cosmic star formation rate. 
Lensed SNe at high redshifts
can in principle also be useful as distance indicators, provided
the magnification is known to high precision. In addition,
strongly lensed, multiply imaged supernovae may also be used to
constrain cosmological parameters through 
the time delay measurements of the SN images \citep{multisne}.

Similar work on the feasibility of galaxy clusters as GTs was
carried out by \citet{sulli}. They focused on data sets through the HST
$I_{814W}$-band with limiting magnitudes between 26.0 and 27.0. We extend this
further by also investigating the $Z$- and
$J$-bands where we find larger effects than in $I$-band. This is not surprising
as the redder filters are sensitive to higher-redshift sources which in turn
are fainter.  
 In addition,
we quantify the limiting search magnitudes and source brightness where GTs
enhance or deplete the discovery rates. Also \citep{gal-yam} conducted
an $I$-band search for SNe in HST fields with galaxy clusters.
Earlier feasibility studies were also carried out by \citet{saini}. They
concentrate their study 
to the lens cluster Abell 2218 where the assumed background sources
were supernovae   
Type Ia and IIL and constrained their analysis to SN searches 
at shorter wavelengths.

Our study thus generalises previous
work and is more applicable to instruments currently being
used for high-$z$ SN searches, e.g.~the $Z$- band search with 
the ACS camera on HST by the GOODS Treasury Team \citep{goods,goods2}, or
future missions such as JWST (formerly NGST).
HST/ACS has a considerable depth but a relatively small field-of-view
compared to  
optical cameras used for ground based SN searches. 
However, the small FOV 
makes a good match to the high amplification region of massive clusters at
the moderate redshifts considered here. 

Throughout the paper we use natural units in which $c=G=1$.
We have also adopted the ``concordance'' cosmology with $\Omega_{\rm M}=0.3,\
\Omega_{\Lambda}=0.7$ and $h=0.7$, where $h=H_0/100\ {\rm km}\  {\rm
s}^{-1}\ \! {\rm Mpc}^{-1}$. All magnitudes are in the Vega system.

\section{The lens equation}
\begin{figure}
\resizebox{\hsize}{!}{\includegraphics{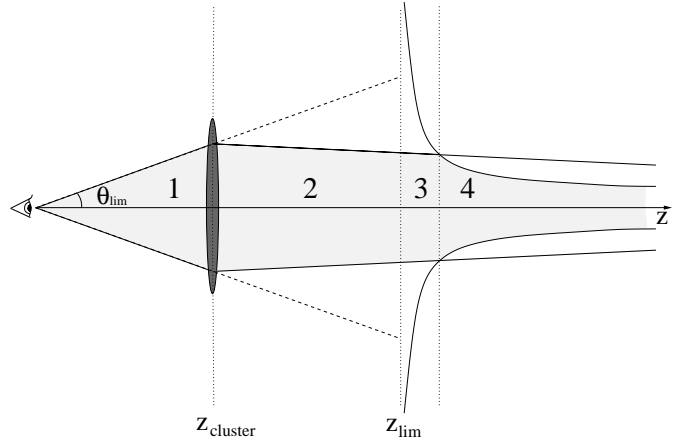}}
\caption{Schematic picture of the lensing situation. $z_{\rm lim}\
(m_{\rm lim})$ 
is the limiting redshift (magnitude) without the lensing cluster. The
shaded area 
indicates the volume where the SNe are bright enough to be 
detected and are visible within the field. Compare the four different
areas 1-4 to the four different
distinguishable $z$-bin ranges of Fig.~\ref{fig:areas}}
\label{fig:sncone}
\end{figure}

In almost all lensing situations of practical astrophysical interest,
the deflection of the light takes place in a very small
fraction of the total light path. This justifies the common
approximation that all deflection occurs in a plane called the
lens plane. So to compute the lensing properties of a halo we 
need to project its mass density onto this plane. This projected mass
density we denote by $\Sigma$. 
Define the 2-D vector $\vek{\xi}$ as the image position (or impact
parameter) in the
lens plane and
$\xi_0$ as an arbitrary length scale in this plane which can be chosen
suitably to simplify the appearance of the equations. Furthermore we
introduce the corresponding quantities in the source plane: $\vek{\eta}$ and
$\eta_0=\xi_0 D_{\rm s}/D_{\rm d}$. $D_{\rm s},\ D_{\rm d}$ and
$D_{\rm ds}$ below are angular diameter distances between observer and
source, observer and lens and lens and source
respectively. $\vek{\eta}$ is the source position.
With these definitions, the lens equation in dimensionless form can be
written as  
\begin{equation} 
\label{lenseq}
\vekt{y}=\vekt{x}-\vek{\alpha}(\vekt{x})
\end{equation}
where $\vekt{x}=\vek{\xi}/\xi_0$, $\vekt{y}=\vek{\eta}/\eta_0$ and
$\vek{\alpha}(\vekt{x})=(D_{\rm d}D_{\rm ds}/\xi_0 D_{\rm
s})\hat{\vek{\alpha}}(\xi_0\vekt{x})$. In this expression
$\hat{\vek{\alpha}}(\xi_0\vekt{x})=\hat{\vek{\alpha}}(\vek{\xi})$ is
the deflection angle at image position $\vek{\xi}$.
We also write the surface mass density in dimensionless form as
$\kappa(\vekt{x})=\Sigma(\xi_0 \vekt{x})/\Sigma_{\rm cr}$, where
$\Sigma_{\rm cr}$ is a critical density given by $D_{\rm s}/4\pi
D_{\rm ds}D_{\rm s}$.
If the projected mass is circularly symmetric (e.g.~for a spherically
symmetric density profile), it is
possible to constrain the impact parameter to the $x^1$-axis in
the lens plane, $\vekt{x}=(x,0)$ if we put
the origin at the center of the lens. This also means that the (scaled)
deflection angle $\vek{\alpha}$ will point towards the lens center and it
will be denoted $\alpha$ instead. 
Due to the nonlinearity of the lens equation, multiple solutions of
Eq.~(\ref{lenseq}) for
$x$ are possible implying more than one image of a single source. 
This occurs when the source,lens and observer are sufficiently
aligned, i.e.~when the source is close enough to the optical axis, see
Fig.~\ref{fig:sncone}. When perfectly aligned, the source will be imaged
as a ring with formally infinite magnification. This ring is called
the (angular) \emph{Einstein radius} (ER) and is usually denoted
$\theta_{\rm E}$. It 
depends on the mass inside this radius but also on the source and lens
redshifts. 
Thus, the Einstein radius defines the area where strong
lensing effects occur.  
A maximum of three images are possible in the models we consider below
and 
the images are called primary, secondary and tertiary, having
different properties which distinguish them. Note that inside
$\theta_{\rm E}$, there are only secondary and tertiary images. For
more details on lensing, see 
e.g.~\citet{schneider}.

\begin{figure}
\resizebox{\hsize}{!}{\includegraphics{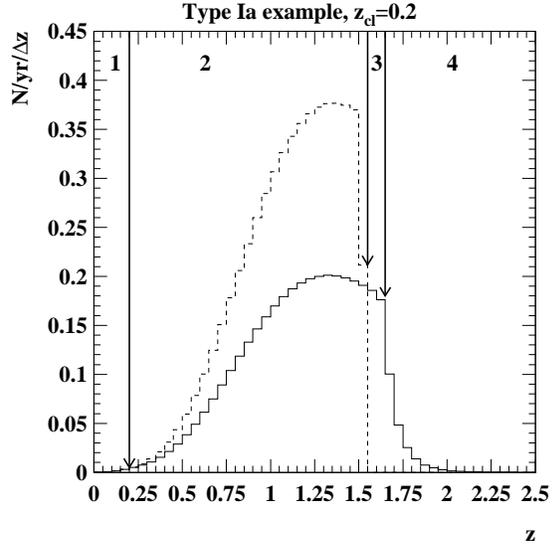}}
\caption{Number of SNe detected per year in redshift bins of $\Delta
z=0.05$. Area 1 at redshifts smaller than the cluster redshift, $z_{\rm cl}$, area 2 between
the lens and $z_{\rm 
lim}$, area 3 where the visible field limits the number of SNe and
area 4 where the magnification is the limitation. The dashed line
shows the number count without the lens and the solid line with a NFW
lens redshift of $z_{\rm cl}=0.2$ and a field-of-view of 16 square arcminutes.}
\label{fig:areas}
\end{figure}

\section{The method}
The number of SNe that can be detected is found by integrating the
number density of supernovae over the volume swept out by the FOV. 
Naturally, we count all SNe in this volume 
that are bright enough to be observed, 
either intrinsically or after having been magnified by lensing, see
Fig.~\ref{fig:sncone}. The trumpet shaped shaded area beyond $z_{\rm lim}$
indicates where the SNe are magnified above the detection
threshold, $m_{\rm lim}$. This figure is to be compared to
Fig.~\ref{fig:areas}, which is an example of a typical set of
parameters for Type Ia 
SNe and a NFW lens and shows the number of detected SNe, $N$, per year
and redshift bin $\Delta z =0.05$. The shaded areas (volumes) denoted 1-4 in
Fig.~\ref{fig:sncone} also have counterparts here. Up to the lens
redshift, $z_{\rm cl}$
(at 0.2 here), there should be no difference between the
cases with and without the lens (area 1). Beyond the lens, there is a
region where the spreading of the field diminishes the area compared
to the case with no lens (area 2), from $z_{\rm cl}$ to $z_{\rm
lim}$. Beyond $z_{\rm lim}$ there is a small area where the visible
field still is the limiting factor to the number count (area
3). Finally, the magnification is limiting in area 4. Under specific
circumstances, e.g.~ if the lens is very massive or at low redshifts
and the FOV is 
small, another effect might become important. As indicated above,
there are no primary images inside the Einstein radius. In all our
considered models, primary images are 
much more 
numerous than secondary or tertiary, and
in principle a
situation can arise where $\theta_{\rm lim}<\theta_{\rm E}$ only
giving secondary and tertiary images at this source redshift. This
will lead to a decrease in the number
count when many sources are affected and the effect can be  
seen in Fig.~\ref{fig:mvirvar}. 

As multiple imaging is possible we have
taken this into account by counting secondary and tertiary images that
are bright enough as
events independent of the primary image, resulting in a slightly
higher rate. As indicated above this is a very small effect in all cases
considered here.

\subsection{Supernova rates, types and magnitudes}

\begin{figure}
\resizebox{\hsize}{!}{\includegraphics{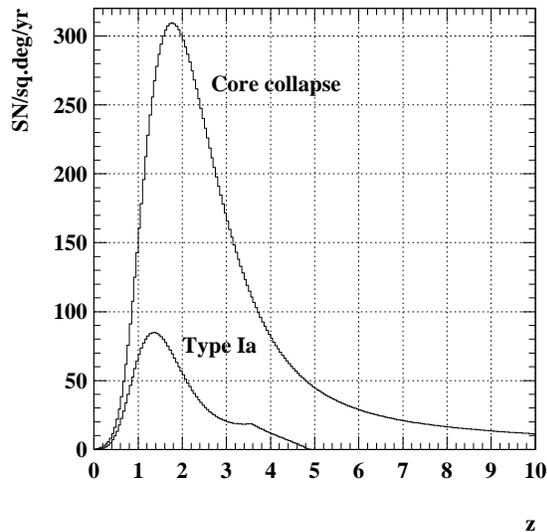}}
\caption{Supernova rates versus redshift in bins of $\Delta
z=0.05$. For core collapse SNe the plot shows
IIn+IIp+IIL+Ibc+87a-like. The increasing time dilation of the SN rate
with redshift is taken into account.The distribution includes neither
lensing nor observational selection effects.}
\label{fig:rates}
\end{figure}

We have concentrated our study on Type Ia and Type II supernovae.
The Type II SNe are divided into three subclasses:
IIn, IIp and IIL, with 
abundances of 0.02, 0.3 and 0.3 rela\-tive to the total core-col\-lapse
(cc) SN rate
(IIn+IIp+IIL+Ibc+87a-like) following the predictions in \citet{dahlfrans}, see
Fig.~\ref{fig:rates}. In order to study the power of galaxy clusters
as GTs we have neglected the intrinsic brightness dispersion of the
supernovae as well as extinction. Thus, the magnitude dispersion considered
in this work stems only from gravitational lensing. For comparison, the
intrinsic properties \citep[from][]{snmag} of SNe are listed in Table
\ref{tab:sn}.  

We have also taken the conservative viewpoint that the SNe are not discovered at
maximum by adding 0.5 magnitudes to the absolute magnitudes in the
table. This has the same effect as decreasing the limiting
magnitude of the telescope by 0.5 magnitudes.

The simple identification of areas 1-4 in Figs.~\ref{fig:sncone} and
\ref{fig:areas} is not possible in the case of our Type II SN
plots. This is due to the fact that they have different luminosities
and will be seen out to different $z_{\rm lim}$ (defined in Fig.~\ref{fig:sncone}). This instead gives
rise to three peaks (at most) in the $N/{\rm yr}/\Delta z$ vs.~$z$
plots for Type II SNe as they show the sum of all three types considered.  

\begin{table}
  \caption{Supernova properties, assuming discovery at maximum
luminosity in $B$-band. Type Ibc and 87a-like events are not considered but are
listed for completeness.}
  \vspace{0.5cm}
    \begin{tabular}{cccc}
    \hline\hline
    Type & Rel.~cc fraction & Abs.~mag. $M$ & Disp.~$\sigma_{\rm M}$ \\
    \hline
    IIn & 0.02 & -18.78 &  0.92 \\
    IIp & 0.3 & -16.61 & 1.23 \\
    IIL & 0.3 & -17.80 & 0.88  \\
    Ibc & 0.23 &  -17.92 &  1.29 \\
    87a-like &  0.15 &  - &  - \\
    Ia & 0 & -19.17 &  0.16\footnotemark \\
    \hline
    \end{tabular}
  \label{tab:sn}
\end{table} 
\footnotetext[1]{After width-brightness correction.}

\subsection{Cluster models}
\label{sec:clumod}

We use two different models for the clusters, the singular isothermal sphere
(SIS)
and the Navarro-Frenk-White profile (NFW)
obtained in N-body simulations \citep{nfw} on which we will focus. The
large lens masses 
considered here seem to be better fitted by the NFW profile 
\citep{liostriker} but we will see in
Sect.~\ref{sec:simres} that they do not differ much for the cases
we consider.
 
There is also an ongoing debate on whether the dark matter halos are
centrally cuspy
or possess a roughly constant density core. The issue is not settled
but there is some evidence
of cuspy central regions on cluster scales which is what we have
assumed in this paper \citep[see e.g.][]{steen}.

For a description of the lensing properties of the SIS, see
e.g.~\citet[Ch.~8.1.4]{schneider}.
The SIS only has primary and secondary images as the tertiary image
always is infinitely faint. 

Gravitational lensing by NFW halos is well described in
\citet{wrightbrain} and we refer to that paper for further reading.   
To find the parameters for a specific cluster mass and redshift we use the
Fortran 77 code {\tt charden.f} available on 
the homepage of Julio Navarro\footnote{http://pinot.phys.uvic.ca/\~{}jfn/mywebpage/jfn\_I.html}.
The NFW profile produces both primary, secondary and tertiary images.

\begin{figure}
\resizebox{\hsize}{!}{\includegraphics{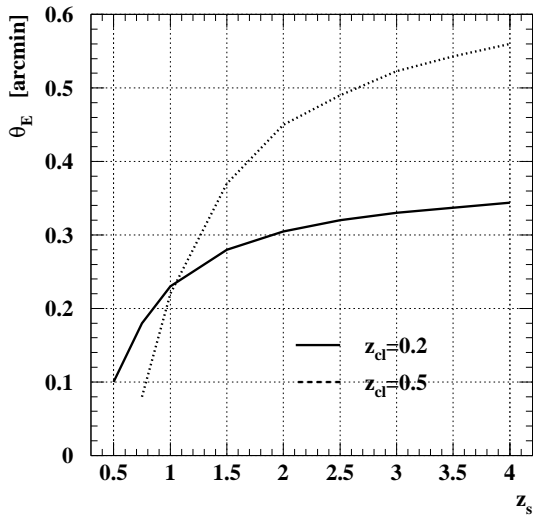}}
 \caption{Redshift dependence of the Einstein radius, within which
 strong lensing effects occur, for two different cluster redshifts,
 $z_{\rm cl}$. The cluster virial mass is $1.4\times 10^{15} h^{-1}$ M$_{\odot}$.}
 \label{fig:einstein}
\end{figure}

To meaningfully compare the results for SIS and NFW we have assumed
$M_{\rm vir}$,
the mass within the virial radius, to be equal for both of them. 
The virial overdensity used to obtain the virial radius is computed
using the results found in \citep{bryan} which is a fit to the
calculation in \citep{eke}. However, the Einstein radii of a NFW and a SIS halo
of equal virial mass will be quite different. For a $1.4\times
10^{15}\ h^{-1}$ M$_{\odot}$ halo at $z_{\rm cl}=0.2$, $\theta_{\rm
E}^{\rm SIS}\sim 2\theta_{\rm
E}^{\rm NFW}$ for source redshifts between 1.5 and 4. This difference
in ER does not
affect the results strongly as is seen e.g.~in Fig.\ref{fig:mlvar0205II}. 
In Figure \ref{fig:einstein} we have plotted the source redshift dependence
of the Einstein radius, within which strong lensing effects occur,
for two different $1.4\times 10^{15} h^{-1}$ M$_{\odot}$ NFW cluster redshifts. As can be seen, the ER has a
weaker dependence on $z_{\rm s}$ for higher redshifts.

\subsection{Telescope and cluster parameters}

The main reason for studying gravitationally magnified supernovae is
that it might enable us to explore supernova explosions otherwise 
too faint to be detected by the search instruments currently used
or being planned. 
Because these SNe are predominantly at very high redshifts we focus
on the use of the largest wavelength optical filters and NIR
for their discovery. We concentrate on the feasibility to discover 
magnified SNe in the $I$-, $Z$- and $J$-bands.


We are assuming a circular 16 sq.~arcminutes field centered on the 
galaxy cluster. The considered solid angle is thus comparable to the
foreseen FOV of NIRCam/JWST and  
slightly larger than the 
dimensions of the Advanced Camera for Surveys (ACS, 3\farcm37
$\times$ 3\farcm37), 
on HST where SN searches are currently conducted in $Z$-band by e.g.~the GOODS 
team\footnote{http://www.stsci.edu/ftp/science/goods/transients.html}.
Ground based NIR cameras on 8-m class telescopes with deep imaging
capabilities  
in $J$-band like ISAAC/VLT (2\farcm5\arcmin $\times$ 2\farcm5), 
NIRI/Gemini (2\arcmin $\times$ 2\arcmin) or CISCO/Subaru (2\arcmin $\times$ 2\arcmin) are also
contained within the maximum FOV considered. We also briefly consider
searches in $I$-band where 
most of the efforts of the SCP and High-Z teams are concentrated, although with significantly
larger FOV. K-corrections were calculated using the SNOC package \citep{snoc}.

\begin{figure}
\resizebox{\hsize}{!}{\includegraphics{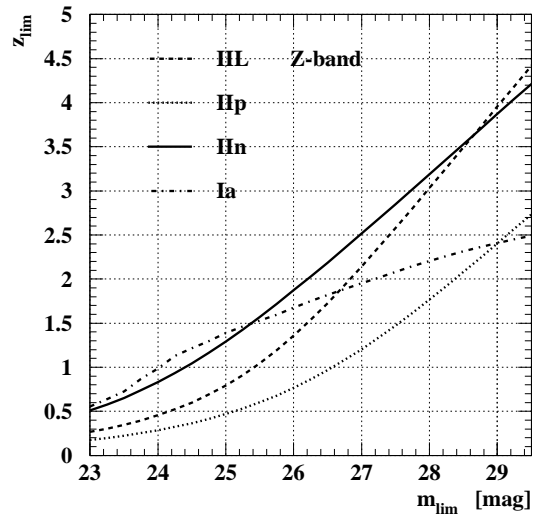}}
 \caption{Redshift reach as a function of limiting ($Z$-band) magnitude in the supernova search. For comparison, the
 ACS/GOODS transient survey has a limiting magnitude around $Z$=25.5.}
 \label{zlim}
\end{figure}

We have investigated limiting magnitudes ranging from 22nd to 30th
magnitude, reaching up
to a limiting redshift $z_{\lim} \sim 4-5$ for some of the SN types considered.
Primarily, we study the properties of quite heavy clusters of virial
mass $1.4\times 10^{15}\ h^{-1}$ 
M$_{\odot}$ as GTs but we also briefly consider a cluster of mass
$1.4\times 10^{14}\ h^{-1}$ M$_{\odot}$. Cluster redshifts between
$z_{\rm cl}=0.2$ and 
$z_{\rm cl}=1.0$ were assumed.

\section{Simulation results}
\label{sec:simres}

\subsection{Supernova searches in $Z$-band}
\label{sec:zband}

First, we consider the case of observational programs such as the ongoing
ACS/GOODS transient 
survey, comparing the potential gain in the discovery rate if the
observations were  
done looking towards a heavy galaxy cluster.  These search
observations are done in $Z$-band, with 
an approximate limiting magnitude of 25.5 mag.

\begin{figure}
\resizebox{\hsize}{!}{\includegraphics{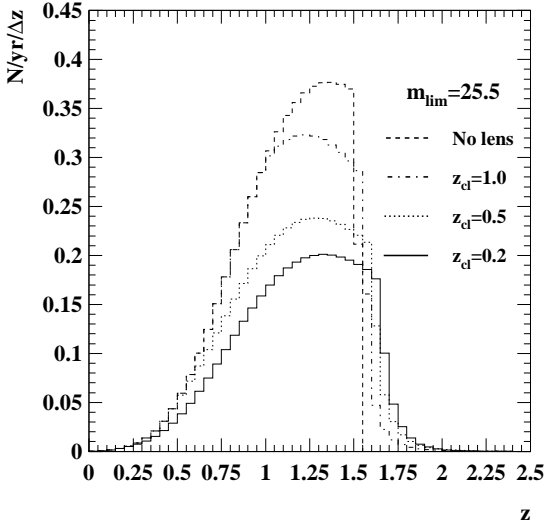}}
 \caption{Number of Type Ia SNe detected per year and $\Delta z=0.05$
 vs.~$z$ in the $Z$-band for various cluster redshifts, $z_{\rm
 cl}$. The FOV is 16 sq.~arcminutes while the limiting magnitude is
chosen to be what is currently used at ACS for the GOODS transient survey.}
 \label{fig:bzIafiltz}
\end{figure}

\begin{figure}
\resizebox{\hsize}{!}{\includegraphics{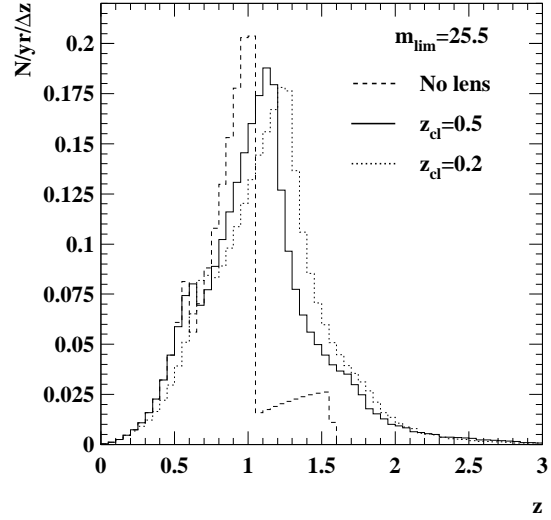}}
 \caption{Number of Type IIn+p+L SNe detected per year and $\Delta
 z=0.05$ vs.~$z$ in the $Z$-band for two different cluster redshifts,
 $z_{\rm cl}$. The FOV is 16 sq.~arcminutes.}  
 \label{fig:bzIIfiltz}
\end{figure}

\begin{figure}
\resizebox{\hsize}{!}{\includegraphics{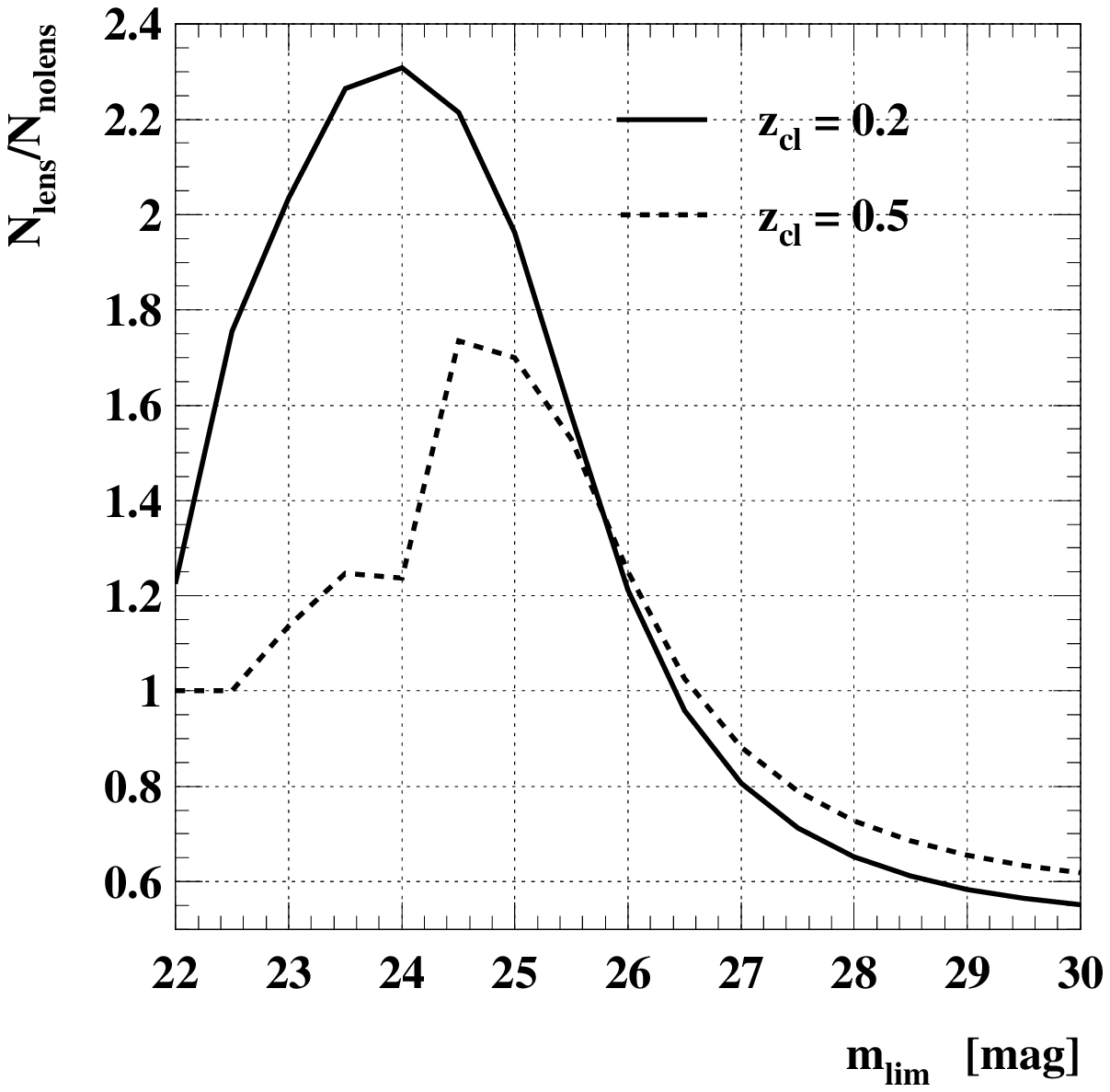}}
 \caption{Gain factor of Type IIn+p+L SNe detected
 vs.~$m_{\rm lim}$ in the $Z$-band for two different cluster redshifts,
 $z_{\rm cl}$. The FOV is 16 sq.~arcminutes.}
 \label{fig:mlvarIIfiltz}
\end{figure}

If nothing else is stated, the lens model we use is a NFW halo. 
Figs.~\ref{fig:bzIafiltz} and \ref{fig:bzIIfiltz} show the 
expected number of discoveries in a 16 sq.~arcminutes field for 
Type Ia and Type II SNe respectively. In the Ia case, although the search
becomes significantly deeper, the effect of spreading the field
dominates, giving 
an overall decrease in the number count, relative to the case with no
lens, of more than 30 \% for the assumed $z$-dependence of the rates. The
qualitative  
effect proves to be roughly independent of the
lens redshift even though it is less severe for higher cluster
redshifts. For the fainter Type II SNe, the effect is reversed at this limiting
magnitude and more supernovae could be detected at redshifts far beyond the
cutoff without the lens. The quantitative gain can be seen in
Fig.~\ref{fig:mlvarIIfiltz} where the gain factor, $N_{\rm
lens}/N_{\rm nolens}$, is shown
as a function of limiting magnitude. At
$m_{\rm lim}=25.5$, a factor of 1.5 to 1.6 is obtained depending on
the cluster redshift. 
Since the simulations were quite
time-consuming, some time was saved by not computing too many points
in each plot. This gives the artificial non-smooth behavior seen in
the figure. The dip
at $m_{\rm lim}=24$ for $z_{\rm cl}=0.5$ originates from the fact that
$z_{\rm lim}(m_{\rm lim}=24)\simeq 0.46$ for Type IIL SNe
i.e.~slightly less than the lens redshift implying that all SNe of this
kind beyond
the cluster are missed in our ``standard candle''
treatment. Increasing $m_{\rm lim}$ 
pushes $z_{\rm lim}$ above 0.5. Thus, a sudden rise is expected since
the Type IIL SNe behind the lens that are sufficiently magnified are
seen. If these SNe were not present, the curve would have followed the
qualitative behavior of the $z_{\rm cl}=0.2$ curve, i.e.~decreasing
after $m_{\rm lim}\sim 24$.  

\subsection{Supernova searches in $J$-band}
Next, we consider the possibility of doing the searches at 
near-IR wavelengths, e.g.~in $J$-band. In this discussion we 
ignore the observational difficulties of such a program, e.g.~the 
rather long exposure times required to overcome 
the high sky noise due to atmospheric emission if a ground based
search were to be attempted. Searching with NICMOS on HST would
be an alternative. A limiting factor is the 
small size of the HgCdTe arrays. 

\begin{figure}
\resizebox{\hsize}{!}{\includegraphics{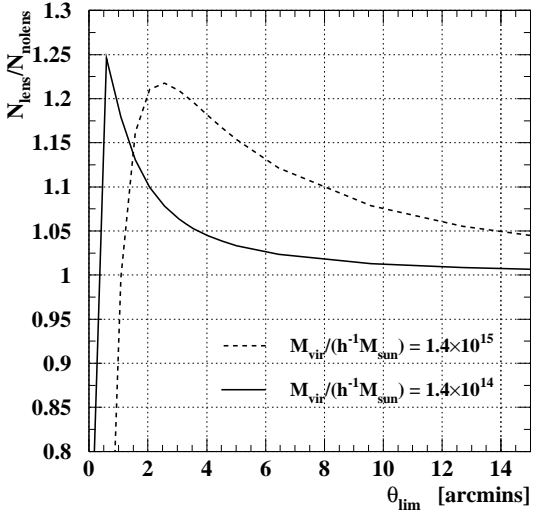}}
 \caption{Gain factor of Type IIn+p+L SNe detected
 vs.~$\theta_{\rm lim}$ for two different cluster masses $M_{\rm
 cl}$ and $z_{\rm cl}=0.2$. The limiting magnitude is 26 for this
 plot, using the $J$-band.}  
 \label{fig:mvirvar}
\end{figure}

To see what effect the FOV and different cluster masses have on the
simulations we plot the gain factor vs.~$\theta_{\rm lim}$
(defined in Fig.~\ref{fig:sncone}) for two different cluster masses
and $z_{\rm cl}=0.2$ in
Fig.~\ref{fig:mvirvar}. 
The aforementioned effect that for a small FOV it is possible to end up
inside the Einstein radius where no primary images of sources at some
specific distance are found can be seen for small $\theta_{\rm lim}$ 
for both clusters. Although $\theta_{\rm E}$ depends on the source
redshift, the dependence is rather weak for the higher redshift
range as already noted. Between $z=1.5$ and $z=4$ the Einstein radius
changes by about 
20\% and is $\sim 0\farcm 3$ for the heavier of the two clusters as
can be seen in Fig.~\ref{fig:einstein}.     

\begin{figure}
\resizebox{\hsize}{!}{\includegraphics{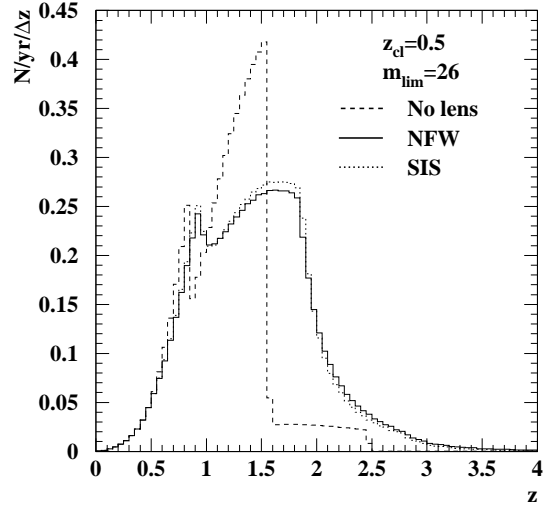}}
 \caption{Number of Type IIn+p+L SNe detected per year and $\Delta z=0.05$
 vs.~$z$ in the $J$-band. The cluster redshift is 0.5 and the FOV is
 16 sq.~arcminutes. This plot
 shows the small difference between 
 the NFW and SIS profiles at this limiting magnitude.}
 \label{fig:bz05m26}
\end{figure}

\begin{figure}
\resizebox{\hsize}{!}{\includegraphics{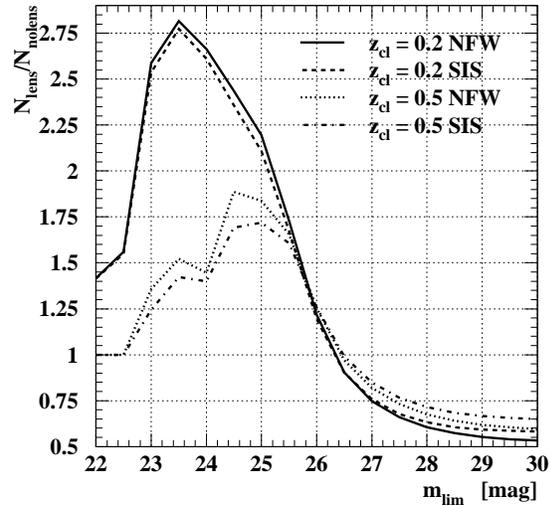}}
 \caption{Gain factor of Type IIn+p+L SNe detected
 vs.~$m_{\rm lim}$ in the $J$-band for two different cluster
 redshifts, $z_{\rm cl}$. The small difference between NFW and SIS is
 also seen in this plot. The FOV is 16 sq.~arcminutes.}
 \label{fig:mlvar0205II}
\end{figure}

\begin{figure}
\resizebox{\hsize}{!}{\includegraphics{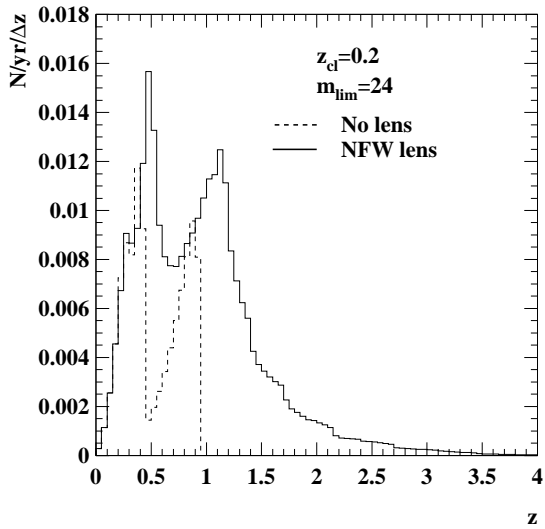}}
 \caption{Number of Type IIn+p+L SNe detected per year and $\Delta
 z=0.05$ vs.~$z$ in the $J$-band for a cluster at $z=0.2$ and with a
 limiting magnitude of 24. The FOV is 16 sq.~arcminutes.}
 \label{fig:bz02m24}
\end{figure}

\begin{figure}
\resizebox{\hsize}{!}{\includegraphics{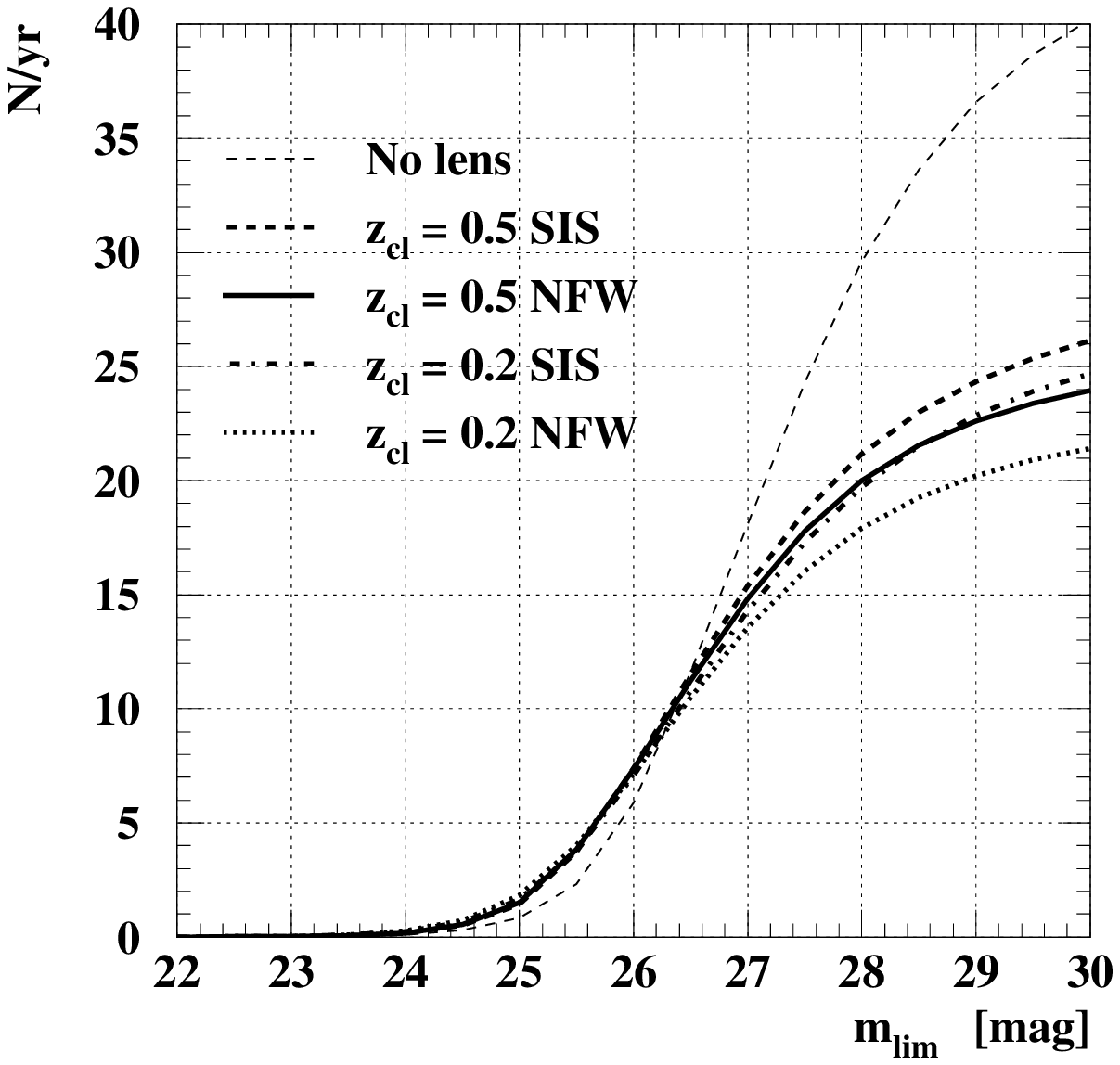}}
 \caption{Number of Type IIn+p+L SNe detected per year vs.~$m_{\rm
 lim}$ in the $J$-band for two different cluster redshifts, $z_{\rm
 cl}$. The FOV is 16 sq.~arcminutes and the distribution is cumulative}
 \label{fig:mlvarabs0205II}
\end{figure}

Figure \ref{fig:bz05m26} 
shows a gain in
the number count discernible by comparing the areas under the
graphs with and without the lens. This is seen quantitatively in
Fig.~\ref{fig:mlvar0205II} as a factor $\sim 1.25$ at $m_{\rm lim}=26$
for both lens redshifts. Figures
\ref{fig:bz05m26} 
and \ref{fig:mlvar0205II} 
also makes it quite clear that
the difference between the NFW and SIS models is small. Again we
see the dip at $m_{\rm lim}=24$ in Fig.~\ref{fig:mlvar0205II}. To continue
the comparison; we see a quite large increase in the gain factor at
$m_{\rm lim}\sim 24$ (max.~at 23.5 to be precise) for the $z_{\rm
cl}=0.2$ cluster, 
and
therefore we display in Fig.~\ref{fig:bz02m24} the distribution of
these Type IIs in redshift space for $m_{\rm lim}=24$ showing a clear
and large displacement of the distribution towards higher redshifts
compared to the case with no lens. This would seem particularly interesting
as $J=24$ is clearly within reach for 8-m class ground based telescopes.
However,
Fig.~\ref{fig:mlvarabs0205II} reveals that the absolute number of SNe
for this limiting magnitude is quite low. One important thing to note
though is that the absolute numbers are uncertain since they depend
strongly on the supernova rates which in turn are quite uncertain, especially
for sources at higher redshifts. 

\begin{figure}
\resizebox{\hsize}{!}{\includegraphics{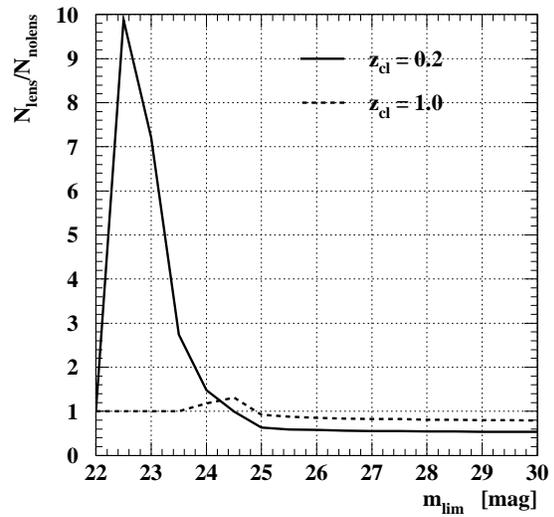}}
 \caption{Gain factor of Type Ia SNe detected vs.~$m_{\rm
 lim}$ in the $J$-band for two different cluster redshifts, $z_{\rm
 cl}$. The FOV is 16 sq.~arcminutes.}
 \label{fig:mlvar0210Ia}
\end{figure}

\begin{figure}
\resizebox{\hsize}{!}{\includegraphics{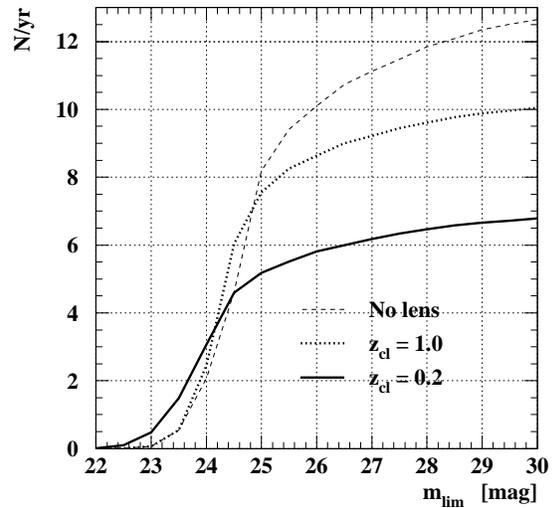}}
 \caption{Number of Type Ia SNe detected per year vs.~$m_{\rm lim}$ for
 two different cluster redshifts, $z_{\rm cl}$. The FOV is 16
 sq.~arcminutes and the distribution is cumulative.}
 \label{fig:mlvarabs0210Ia}
\end{figure}

If we now study similar plots for Type Ia SNe we see a huge increase
in the gain factor at small $m_{\rm lim}$
$\sim$22-23 in Fig.~\ref{fig:mlvar0210Ia} for the $z_{\rm cl}=0.2$
cluster. Unfortunately this is not so 
exciting when comparing with Fig.~\ref{fig:mlvarabs0210Ia} where
we again see that the absolute numbers are quite low. However, in the
region \mbox{$23.5\lesssim m_{\rm lim}\lesssim 24.5$} the situation looks a
bit more interesting for the nearer cluster. 

\subsection{Supernova searches in $I$-band}
\begin{figure}
\resizebox{\hsize}{!}{\includegraphics{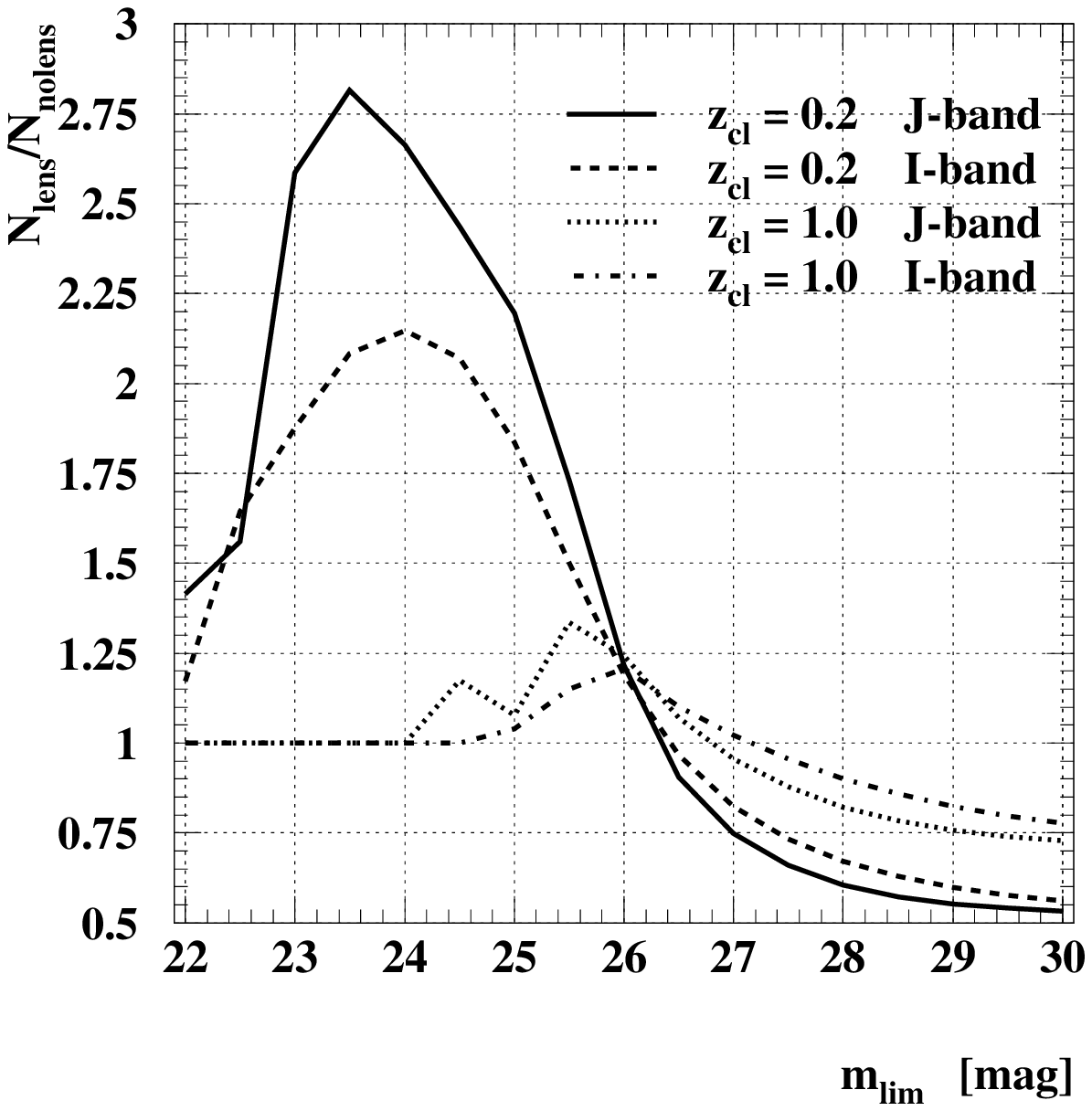}}
 \caption{Gain factor of Type IIn+p+L SNe detected
 vs.~$m_{\rm lim}$. in the $I$- and $J$-bands for two different
 cluster redshifts, $z_{\rm cl}$. The FOV is 16 sq.~arcminutes.}
 \label{fig:mlvarIJtII}
\end{figure}

Finally, we briefly compare the results shown so far 
with the expectations from an $I$-band SN search, the currently
favoured search filter 
for ground based high-$z$ supernova searches.
Fig.~\ref{fig:mlvarIJtII} shows the expected benefit of heavy clusters as
GTs at two redshifts for $I$- and
$J$-band. The gain in the $I$-band is generally smaller
in the interesting region where the gain is greater than one except
maybe for a high redshift cluster. However, clusters this heavy at
such high redshifts are very rare \citep{rosati}. Again, the bumpy
behaviour of the $z=1.0$ $J$-band graph is due to the effect described
in Sect.~\ref{sec:zband} for Type IIL SNe. 

\section{Discussion}
As stated above we have treated all SNe as standard candles. This
perhaps crude assumption when applied to Type II SNe might appear plausible
because one can assume approximately equally many SNe brighter than average
as fainter for each SN type. This could then be assumed to average
out. However, the effect of adding a dispersion is not trivial
since it will affect the lensed and the unlensed case differently. The
magnification will boost the faint end tail,
thereby adding from this tail supernovae in area 2 of
Fig.~\ref{fig:sncone} to the number count. The effect on the unlensed
case will be to add some SNe beyond and remove some SNe
prior to $z_{\rm lim}$ but with a larger volume for the ones beyond. 
The effect will also depend on the differential SN rate, especially
around $z_{\rm lim}$. A more careful investigation should of course
include the SN brightness dispersion.  

\begin{figure}
\resizebox{\hsize}{!}{\includegraphics{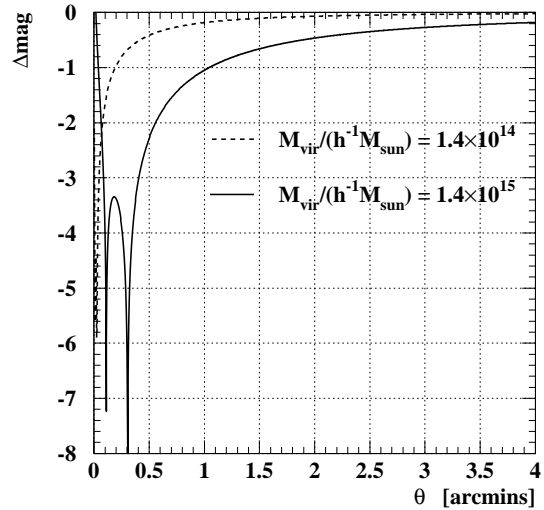}}
 \caption{Magnification as a function of the impact parameter in the
 lens plane for a source at $z=2$. The lens redshift is 0.2.}
 \label{fig:magnplot}
\end{figure}

A more robust justification is found by studying the
magnification as a function of the impact parameter, see
Fig.~\ref{fig:magnplot}. For a $1.4\times 10^{15}\ h^{-1}$ M$_{\odot}$
NFW cluster 
at $z_{\rm cl}=0.2$ and a 
source at $z=2.0$ it is seen that within 1$\arcmin$ of the cluster
core, the magnification exceeds one magnitude, approximately the intrinsic
dispersion of the Type II SNe considered, see Table \ref{tab:sn}. Thus,
within this radius the 
magnification dominates the magnitude dispersion for this
lens-source configuration. The situation does not change much for
sources in the most interesting regions where the SN rate is
high. However for lighter clusters we see that the approximation
becomes worse. 
Similarly, extinction losses have been neglected. Thus, the net benefit
of gravitational lensing is somewhat underestimated, as GTs would
certainly enhance the detectability of extincted supernovae.

In Fig.~\ref{fig:magnplot} we can also identify the regions where
secondary and tertiary images appear. Tertiary images show up in the
innermost regions up to the first big dip. Secondary images are
located between the two dips and outside the second dip (the locus of
the Einstein radius) there are only
primary images. Although not obvious from Fig.~\ref{fig:magnplot},
in all our considered models primary images are 
much more abundant for larger fields-of-view and roughly
make up about
95 \% of the images. However, the fraction of secondary and tertiary
images is larger for lower redshift clusters (and also for SIS clusters which
are more efficient image-splitters) since we then look at the more
central parts of the cluster.        

We have only studied spherically symmetric clusters which of course is
an oversimplification since clusters undoubtedly have substructure in
the form of cluster member galaxies. To see in more detail how a
specific cluster can affect the searches, a better way is to follow the example
in \citet{saini} and use a more sophisticated lens
model.    

The adopted cluster masses in this work, $\sim 10^{15}\ h^{-1}$ M$_{\odot}$,
are consistent with virial and lensing mass estimates of X-ray luminous clusters at 
$z \sim 0.2-0.3$ \citep[see][and references therein]{irgens}.

Finally, since we have left out Type Ibc (and 87a-like) SNe we have
underestimated the total number of SNe that would be detectable.

\section{Conclusions}
Observations pointing behind massive heavy galaxy clusters 
at redshifts $z=0.2-0.5$ may be a useful way
to enhance the detectability of very high-$z$ supernovae.
For the considered supernova rates, the most interesting
effects show up for core-collapse SNe in   
searches at limiting magnitudes  
$m_{\rm lim}\sim 25-26.5$ mag where
the detection rate 
could be significantly enhanced and where the absolute number of
detectable events is considerable. However, the SN rates at high-$z$
and thus also the 
absolute numbers are quite uncertain. The relative gain, which is less
sensitive to the SN rates, peaks at lower
limiting magnitudes of $\sim$24 where the number of detections could
increase by up to a factor 3 and by a much 
larger factor at the highest redshifts. For programs
such as the GOODS/ACS transient survey, the discovery of supernovae 
beyond $z\sim 2$ may be significantly increased with the aid of
GTs. This technique may be used to probe the cosmic star formation 
rate at very high redshifts where there is significant controversy
\citep[see e.g.][]{lanzetta} although this requires good knowledge of
the cluster 
properties such as mass and density profile. 
When interpreting the measured rates of distant SNe as well as their apparent 
brightness, special care must be taken so that magnification effects
from GTs are 
properly accounted for. The potential bias increases with the central
wavelength  
of the filter used for the supernova search, due to the enhancement of 
acceptance for very redshifted sources.    
For extremely deep
observations $m_{\rm lim} > 27$ mag or for very bright SNe (e.g.~Type Ia) the 
competing effect of field reduction dominates and fewer supernovae
are likely to be discovered behind foreground clusters. 
However, the most important effect is that
the detection rate at high redshift could be significantly enhanced
when there is an intervening cluster acting as a gravitational telescope.

\begin{acknowledgements}
The authors would like to thank Anne Green, Edvard M\"ortsell and
Joakim Edsj\"o for helpful
comments and suggestions, H\aa kon Dahle for cluster information and
Julio Navarro 
for providing the NFW code on his homepage.  
CG would like to thank the Swedish Research Council for financial support.
AG is a Royal Swedish Academy Research Fellow
supported by a grant from the Knut and Alice Wallenberg Foundation.
\end{acknowledgements}

\end{document}